**RESEARCH**

# Spatial Distribution of Solar PV Deployment: An Application of the Region-Based Convolutional Neural Network

Serena Y. Kim[1,2,3*], Koushik Ganesan[4], Crystal Soderman[2] and Raven O'Rourke[5,6]

*Correspondence:
serena.kim@ucdenver.edu
[1]College of Engineering, Design and Computing, University of Colorado Denver, 1200 Larimer St., Denver, CO USA, 80204
[2]School of Public Affairs, University of Colorado Denver, 1380 Lawrence St., Denver, CO USA, 80204
[3]Computer Science, University of Colorado Boulder, 1111 Engineering Dr., Boulder, CO USA, 80309
Full list of author information is available at the end of the article

**Abstract**

This paper presents a comprehensive analysis of the social and environmental determinants of solar photovoltaic (PV) deployment rates in Colorado, USA. Using 652,795 satellite imagery and computer vision frameworks based on a convolutional neural network, we estimated the proportion of households with solar PV systems and the roof areas covered by solar panels. At the census block group level, 7% of Coloradan households have a rooftop PV system, and 2.5% of roof areas in Colorado are covered by solar panels as of 2021. Our machine learning models predict solar PV deployment based on 43 natural and social characteristics of neighborhoods. Using four algorithms (Random Forest, CATBoost, LightGBM, XGBoost), we find that the share of Democratic party votes, hail risks, strong wind risks, median home value, and solar PV permitting timelines are the most important predictors of solar PV count per household. In addition to the size of the houses, PV-to-roof area ratio is highly dependent on solar PV permitting timelines, proportion of renters and multifamily housing, and winter weather risks. We also find racial and ethnic disparities in rooftop solar deployment. The average marginal effects of median household income on solar deployment are lower in communities with a greater proportion of African American and Hispanic residents and are higher in communities with a greater proportion of White and Asian residents. In the ongoing energy transition, knowing the key predictors of solar deployment can better inform business and policy decision making for more efficient and equitable grid infrastructure investment and distributed energy resource management.

## 1 Introduction

Over the last decade, solar energy technology has advanced rapidly to the point where the 20-year Levelized Cost of Energy (LCOE) for solar is now lower than that of coal in many countries around the world [3]. However, the average payback period of a solar project is 8 to 10 years for the residential sector as of 2022, meaning the high up-front costs are still significant barriers to investing in solar panels for low- and moderate-income (LMI) households. Not surprisingly, solar adopters still have higher income than non-adopters in 2021 [7]. While more affluent communities have taken advantage of government subsidies and have actively incorporated solar energy into their energy portfolio, LMI communities have lagged in enjoying the benefits of solar energy while remaining more vulnerable to energy costs.

The growth in rooftop solar photovoltaic (PV) deployment has occurred unevenly across neighborhoods with different natural and socioeconomic characteristics. The overall rate of solar energy deployment is lower in racially diverse communities [40] and lower-income neighborhoods [33]. Areas with greater solar radiation [55] and government incentives for renewable energy [43], such as renewable energy portfolio standards (RPS) and net metering (NEM), have higher levels of solar PV adoption. Yet, the mechanisms by which socioeconomic and environmental features collectively shape uneven rooftop solar deployments remains unclear.

This study models and predicts solar PV deployment at the US census block group level using satellite imagery from the state of Colorado, USA in 2021. Using Geographical Information Systems (GIS), we aggregated 43 layers of natural/built environment, social, economic, and policy data. We build a machine learning (ML) model predicting two measures of solar PV deployment, solar PV count per household and PV-to-roof ratio, based on the 43 input features. The state of Colorado provides an apt setting to investigate solar PV deployment because the state has high solar radiation, a variety of natural landscapes, and diverse socioeconomic characteristics across neighborhoods within the state with the total area of 269,837 $km^2$.

This study advances the current knowledge of solar PV deployment in four ways. First, this study is one of the first attempts to measure the proportion of roofs covered by solar panels (henceforth 'PV-to-roof ratio'). Most existing approaches measure solar PV deployment by the number of solar system counts per capita/household [44] or the size of solar panels [55]. The PV-to-roof ratio can provide useful information to utility infrastructure planning and energy policy design as the ratio indicates the availability of underutilized roof areas for future PV installations. Communities that wish to enhance their distributed solar energy generation capacity may consider deploying more solar panels in areas with a low PV-to-roof ratio. In addition, aggregated PV areas can have implications for utility infrastructure planning as utilities may need to adjust utility procurement and upgrade grid infrastructure to ensure the reliability of the electrical grid system.

Second, this study provides one of the finest-scale solar deployment datasets to date. Existing studies on neighborhood-level solar deployment usually use US Census tract-level (optimum size of 4,000 residents per tract) or zip code-level data (8,000 residents per zip code on average), which may be inadequate to capture spatial disparities in PV deployment. Neighborhoods within a tract or a zipcode can have widely varying characteristics in terms of residents' socioeconomic status and living arrangements. In this study, we estimate solar deployment at the US Census block group level (average size of 1,500 residents). Our data are more likely to provide targeted results as the variation of neighborhood-level measures tends to increase with a decrease in the scale of the spatial unit [21].

Third, to the best of our knowledge, this study is the first attempt to examine how natural disaster vulnerabilities relate to residents' decision to deploy rooftop solar. The types of natural disasters may affect people's willingness differently. For example, people in areas with frequent power outages due to extreme cold or heat may be more interested in investing in rooftop solar with storage systems to keep the power on during an outage. Residents who experience frequent hail, hurricanes, or tornadoes may be more hesitant to invest in rooftop solar due to concerns over the potential damage to solar panels. In this study, we examine the relationship between six of the most common natural disasters in Colorado and rooftop solar deployment.

Fourth, leveraging the recent development in data science, our model including 43 predictors provides one of the most comprehensive predictive models of rooftop solar deployment to date. Our model explains about 70% of the variation in PV deployment. Existing solar deployment prediction models may be able to enhance their model performance by including the variables found to be important in our analysis, such as political ideology, types and frequency of natural disasters, and local government rooftop solar permitting rules.

The remainder of this paper is organized as follows. Section 2 provides an overview of related work underpinning the empirical analysis of our paper; Section 3 describes data sources, GIS data processing, computer vision methods, and ML models that are used in this study; Section 4 presents the results and identifies the important predictors of rooftop solar deployment; and Section 5 discusses the implications of this study and concludes with future research agendas.

## 2 Related Work

Existing studies on solar PV deployment often take one of the following approaches: (i) individual or household surveys (e.g., Korcaj et al. [30], Schelly and Letzelter [45]), (ii) datasets collected by governments and utility service providers who manage interconnection processes, or the processes by which rooftop solar is allowed to connect to the grid (e.g., Graziano and Gillingham [20]), or (iii) satellite images that can be used to detect the existence and size of rooftop solar PVs using computer vision models (e.g., Yu et al. [55]). Although the first approach involving surveys allows individual-level targeted data collection, this approach can be costly and time consuming to obtain a geographically representative and comprehensive dataset. The second approach provides accurate data of solar adopters, but address-level solar deployment or interconnection data are usually protected and difficult to obtain for research purposes. Therefore, we chose the third approach. Although the third approach requires more time and computing resources for collecting and processing large satellite imagery datasets, this approach allows us to obtain uncensored and comprehensive data on spatial distribution of solar deployment.

There are a few rooftop solar deployment datasets created using satellite images. Google Project Sunroof [32] estimates the number and the sizes of solar PV installations using satellite images since 2013 to date, but about 25% of areas, mostly rural areas, are not included in their dataset. DeepSolar [55] developed a comprehensive database of solar installations for the contiguous United States, but data are from 2017 or earlier and are inadequate to capture the existing solar deployment which has had a sharp increase for the last five years. The National Renewable Energy Laboratory's (NREL) Distributed Generation Market Demand (dGEN) [47] model has up-to-date solar deployment data but the most granular spatial resolution available to the public is county-level data, which is inadequate to capture neighborhood-level spatial disparities. Thus, our solar deployment dataset was created from scratch, collecting satellite images of Colorado, which is our area of interest.

The predictors of solar PV deployment have been widely studied in the renewable energy literature. Previous studies have found that solar PV deployment positively correlates to residents' income [55], education [46], and age [45]. Black- and Hispanic-majority census tracts have installed less rooftop PV compared with tracts that have no racial/ethnic majority, even after controlling for median household (HH) income [48]. Solar radiation positively predicts PV deployment [13, 55]. More rooftop solar PVs exist in areas with a smaller proportion of renter-occupied homes [20], higher median home value [41], and newer buildings. Existing literature from around the world finds an urban-rural divide in solar PV deployment, but this divide moves in different directions depending on the study context. While rural municipalities have more solar PV projects per capita in Switzerland [50], urban counties have more solar installations per capita in Georgia, USA [51]. Recent research has found that solar deployment occurs in many Republican households, but solar deployment continues to occur to a greater extent in Democratic households [36]. Disadvantaged communities that suffer from socioeconomic, health, and environmental burdens are significantly less likely to deploy solar than their more advantaged counterparts, even after controlling for median household income [33].

We identify four important gaps in the literature on the predictors of solar PV deployment. First, the relationship between tree canopy cover and residents' decisions for PV deployment has not yet been examined. Second, although the proximity to transmission lines can affect the performance of distributed solar PVs, no studies have investigated the existence or the size of transmission infrastructure on building owners' decisions to deploy solar. Third, although existing literature has examined how solar PV deployment relates to state-level policies, little is known about the impact of local-level policies and rules such as solar mandates and Solar Permitting, Inspection, and Interconnection (PII) rules. Fourth, there is little empirical research on how the types and intensity of natural disasters influence solar PV adoption and the size of deployment.

## 3 Data and Materials

### 3.1 Satellite Imagery Data Collection

To specify the geographical areas for data collection, we excluded census blocks that do not have any residents because we are only interested in rooftop solar, not the large-scale ground-mounted solar. Of 201,062 blocks in Colorado, 55,258 blocks with zero residents were excluded from data collection. Figure 1 shows 145,804 included blocks.

We developed two measures of rooftop solar deployment: (i) the number of rooftop solar panels per household (PV count per HH) and (ii) the area of solar panels to roofs (PV-to-roof area ratio). To obtain images including all roofs in Colorado, we used Google Maps JavaScript API for downloading Google Earth satellite images. Census block polygons were split into smaller polygons to obtain high resolution images. To that end, we obtained 652,795 image tiles as 640 × 640 pixels at zoom 20. The latitude and longitude of each polygon's center and boundary were used to retrieve the images corresponding to the census blocks. We used World Geodetic System WGS84 format for retrieving the spatial information of the polygons. We collected satellite images from January 2021 through February 2021. The images

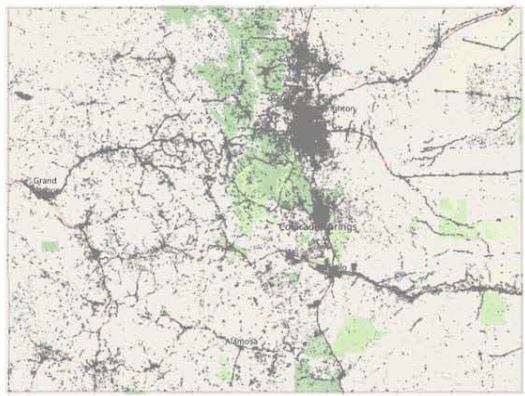

Figure 1: Gray areas are the selected areas for satellite imagery collection

in Colorado were captured by Google Maps between September 2020 and February 2021.

### 3.2 Satellite Imagery Data Processing: Faster RCNN

To identify solar panels and roofs from satellite images, we used the Faster RCNN ( Regions with Convolutional Neural Networks) model [42]. The Faster RCNN model has been one of the widely used state-of-the-art object detectors that outperforms YOLO, SSD, and other traditional models on several key metrics [4, 19]. The essential feature that distinguishes Faster RCNN from its predecessors is the Region Proposal Network (RPN) which significantly reduces computing time and improves performance as it shares layers with the detection stages. For the backbone network, we used ResNet-50, pre-trained on COCO train2017. We also replaced the pre-trained head with a FastRCNNPredictor.

To improve the performance of image detection, we created two copies of the pre-trained Faster-RCNN model, one to identify roofs and the other to identify PVs. We trained our model on 367, 480 × 480 sized, annotated images (image and its corresponding XML file containing bounding box coordinates along with class) by constructing train (80%), development (10%), and test (10%) sets. An example of such an annotated image is shown in Figure 2.

We used the TorchVision library to import the model and the AdamW optimizer [29] to minimize cross-entropy loss. To fine-tune the other hyper-parameters we performed a number of experimental tests on the development set. For the final model, we used a learning rate of $2 \times 10^{-4}$, a weight decay of 0.001. We also used a StepLR as our scheduler with decaying the learning every epoch by gamma of 0.7 along with a batch size of 8 and trained for 4 epochs on Tesla K10 GPUs. After obtaining the prediction, we use non-maximum suppression (NMS) with an Intersection over Union (IoU) of 0.2 and 0.1 cutoffs for roofs and PVs, respectively, over the predicted frames in order to obtain the best model. As a performance measure for determining the accuracy of prediction, we used the mean Average Precision (mAP) for an IoU with the minimum threshold of 0.5 for all object sizes. The IoU is given by

$$\text{IoU} = \frac{\text{area}(A_p \cap A_{gt})}{\text{area}(A_p \cup A_{gt})} \quad (1)$$

where $A_p$ is the predicted frame and $A_{gt}$ is the ground truth frame. Our model achieved an mAP of .95 and .81 for detecting roofs and PVs, respectively.

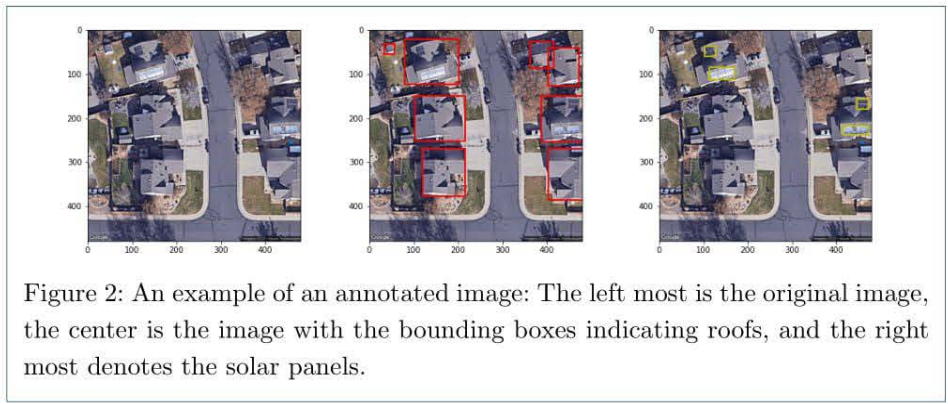

Figure 2: An example of an annotated image: The left most is the original image, the center is the image with the bounding boxes indicating roofs, and the right most denotes the solar panels.

### 3.3 Two Measures of Rooftop Solar Deployment

We created two measures, *PV Count Per Household* and *PV-to-Roof Area Ratio*, of rooftop solar PV deployment by running our Faster RCNN model on 652,795 satellite imagery data.

$$\text{PV Count Per Household (HH)} = \sum_{n=1}^{k} \frac{\text{Number of Solar PV Systems}_k}{\text{Number of Households}_k} \quad (2)$$

$$\text{PV-to-Roof Area Ratio} = \sum_{n=1}^{k} \frac{\text{Solar PV Area}_k}{\text{Roof Area}_k} \quad (3)$$

where $k$ is the number of images for a geographic area (i.e., block group). On average, 189 images correspond to a block group.

### 3.4 Predictors of Solar PV Deployment

The second objective of our study is to understand the predictors of rooftop solar deployment. To build a comprehensive model of rooftop solar deployment, we considered four groups of predictors: (1) natural environment, (2) demographics and built environment, (3) energy infrastructure, market, and policy, and (4) social and natural disaster vulnerabilities.

#### 3.4.1 Natural Environment

***Solar Radiation.*** Solar radiation data are from the national solar radiation database (NSRDB). We used Direct Normal Irradiance (DNI), which is the amount of solar energy received per unit area by a surface that is always held perpendicular to the rays. The DNI measure used in the analysis is obtained by averaging maximum and minimum DNI measured at the centroid of each block, and then the block-level measures are averaged over all blocks within a block group.

***Tree Canopy Cover.*** Using the tree canopy cover data from the United State Forest Services (USFS), we created the % of tree-to-land area variable, which measures the size of the areas covered by trees over the land size in each block group.

*3.4.2 Demographics and Built Environment*

***Demographics.*** Median household income, race and ethnicity (i.e., % White, % Hispanic, % African American, % Asian, % Other race), median age, the proportion of persons with bachelor's or higher, and the proportion of households with at least one person older than 65 years old. Data are from the 2019 American Community Survey (ACS).

***Housing Characteristics.*** Using the ACS block group-level data, which is the most granular spatial resolution available for most housing characteristics, our model includes the proportion of renter-occupied housing units, median home values, and the median year of structures built. Approximately 2% of the missing estimates for the median home values and median year of structures built are imputed with the average values of the adjacent block groups.

***Rural-Urban Classification.*** To account for the differences between rural and urban areas, rural-urban continuum codes from the U.S. Department of Agriculture (USDA) [38] were included in our analysis. The codes range from 1 to 9 where higher values indicate increasing rurality.

***Political Ideology.*** The percentage of votes for the 2020 Democratic presidential candidate at the county level is included in our models to account for heterogeneity in the political orientation of residents. Data are from the MIT Election Data and Science Lab [15].

*3.4.3 Energy Infrastructure, Market, and Policy*

***Transmission Lines.*** Building owners' decision to install rooftop solar could be influenced by the existence of large grid infrastructure such as a high-power electrical transmission tower in their neighborhood. Thus, we included transmission data from Homeland Infrastructure Foundation-Level Data (HIFLD) by aggregating the length and voltage of the transmission lines within each block group level.

***Utility Ownership.*** Since utilities generally develop and implement interconnection standards which define how rooftop PVs can connect to the grid, our analysis includes three types of utility ownership: investor-owned utilities (IOUs); Municipal utilities (MOUs); and rural electric cooperatives (co-ops). Utility ownership data are obtained by spatially merging the map of block groups and the electric utility territory map.

***Electricity Price.*** Using zip-code level utility rate data from the Utility Rate Database (URDB) [1], we obtained residential, commercial, and industrial utility rates for each block group. When a block group has multiple zip codes, the most populous zip code within the block group was used to extract the utility rates for the block group.

***Solar Mandates.*** Solar mandates, a building code that requires new construction homes to be solar-ready or to have a PV system installed, can address the cost-prohibitive barriers of retrofitting a roof or removing shade obstructions associated with solar PV adoption [53]. We obtained local-level solar mandate information from municipal building codes publicly available on municipality web pages.

***Solar Permitting, Inspection, and Interconnection (PII) rules.*** Local (i.e., city, town, county) PII requirements can affect the duration of solar PV installations significantly [39]. Thus, our models incorporate four PII rule variables: (i) SolSmart Awardee (1 for a local jurisdiction awarded by the SolSmart award for improvements to local permitting, inspection, planning, zoning, and/or market development to facilitate solar installs and mitigate associated soft costs, or 0 otherwise), (ii) Online Permit (1 for a local jurisdiction accepting permit submissions through an online portal or email, or 0 otherwise), (iii) Same-day In-person Permit (1 for a local jurisdiction offering over-the-counter permit submission and approval, or 0 otherwise), and (iv) Permit & Pre-Install Days (median business days between first permit submission to the local jurisdiction and approval). Data are from NREL's SolarTrace [12].

3.4.4 Social and Natural Disaster Vulnerabilities

**Social Vulnerability** Utilizing the Social Vulnerability Index (SVI) from the Centers for Disease Control and Prevention (CDC) [18], we specifically consider nine social vulnerability variables: (i) % Below Poverty (percentage of persons below poverty), (ii) % Disability (percentage of civilian non-institutionalized population with a disability), (iii) % Single Parent (percentage of single parent households with children under 18), (iv) % Limited English (percentage of persons (age 5+) who speak English "less than well"), (v) % 10+ Unit Housing (percentage of housing in structures with 10 or more units), (vi) % Mobile Home (percentage of mobile homes), (vii) % People > Rooms (percentage of occupied housing units with more people than bedrooms), (viii) % No Vehicles (percentage of households with no vehicle), (ix) % Unemployed (unemployment rate). We merged our block-group data with the census-tract level SVI data which is the most granular spatial resolution available.

**Natural Disaster Vulnerability** The resilience value of rooftop solar has recently gained more attention. Rooftop solar, combined with batteries, has the potential to provide electricity during power outages caused by extreme weather events. However, certain types of natural disasters, such as frequent hails or tornadoes, can instead discourage investment in rooftop solar because they can damage roofs and PV systems. Therefore, we expect different types of natural disasters differently impact PV deployments. Of 18 natural hazards included in National Risk Index from Federal Emergency Management Agency (FEMA) [2], we incorporated the expected annual loss scores (EALS) of six of the most frequent natural disasters in Colorado. These include drought, wildfire, hail, winter weather, strong wind, and tornado risks.

These four groups of determinants – (i) natural environment, (ii) demographics and built environment, (iii) energy infrastructure, market, and policy, and (iv) social and natural disaster vulnerabilities – were merged with the two measures of rooftop solar deployment at the block group level. Figure 3 provides a snapshot of all satellite and geospatial data processing workflow. The energy infrastructure, market, and policy features that correspond to the boundaries of local jurisdictions, utility service areas, or zip code tabulated-areas are spatially merged with the block group map using the GeoPandas Python library [26]. If a block group crosses multiple jurisdictional boundaries (less than 5% of all block groups), the jurisdiction that has the largest share of the area was selected to represent the block group. Descriptive statistics are presented in Appendix A.

3.5 ML Models

To explain and predict rooftop solar PV deployment, we used four ML algorithms: Random Forest [9], CATBoost [17], LightGBM [27], and XGBoost [11]. Four ML models are estimated on four datasets: (i) PV count dataset without energy policy variables (n= 3441), (ii) PV count dataset with energy policy variables (n= 2328), (iii) PV-to-roof area dataset without energy policy variables (n= 3441), (iv) PV-to-roof area dataset with energy policy variables (n= 2328), yielding a total of 16 models. We separated models without policy variables because policy variables were not available for all cities and towns in Colorado, and some block groups are in unincorporated communities, which are not considered to be municipal areas of their own accord.

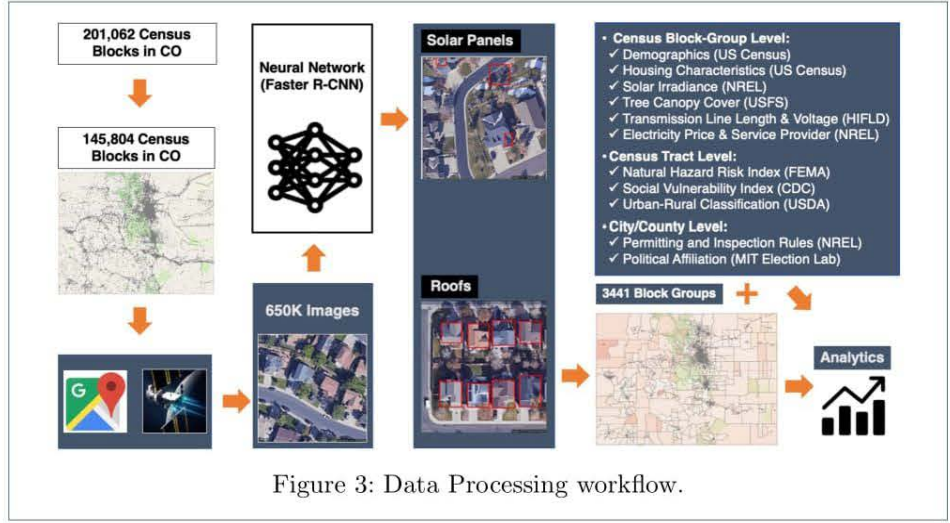

Figure 3: Data Processing workflow.

***Metrics.*** We evaluate the performance of all models based on three measures: Root Mean Squared Error (RMSE), Mean Absolute Error (MAE), and the coefficient of determination ($R^2$).

$$\text{RMSE} = \sqrt{\frac{1}{n} \sum_{i=1}^{n} (y_i - \hat{y}_i)^2} \tag{4}$$

$$\text{MAE} = \frac{1}{n} \sum_{i=1}^{n} \mid y_i - \hat{y}_i \mid \tag{5}$$

$$R^2 = 1 - \frac{\sum_{i=1}^{n} (y_i - \hat{y}_i)^2}{\sum_{i=1}^{n} (y_i - \bar{y})^2} \tag{6}$$

where $n$ is the number of sample (observations) in the dataset, $y_i$ is observed (true) value of the target variable, $\hat{y}_i$ is the estimated (predicted) value of the target variable, and $\bar{y}$ is the mean value of the target variable. The optimization of hyperparameters are provided in Table A1 for each algorithm.

# 4 Results

## 4.1 Rooftop Solar PV Deployment in Colorado

At the block group level (n=3441), 7% of households appear to have a solar panel on their roofs, and 2.5% of roofs are covered by solar panels on average (Figure 4). PV count per household varies from 0% to 78.4%, and the PV-to-roof area ratio varies from 0% to 26%. Any solar panels detected on an image that does not include roofs are excluded from estimation because this study is interested only in rooftop solar PV deployment on buildings, not ground-mounted and utility-scale solar systems.

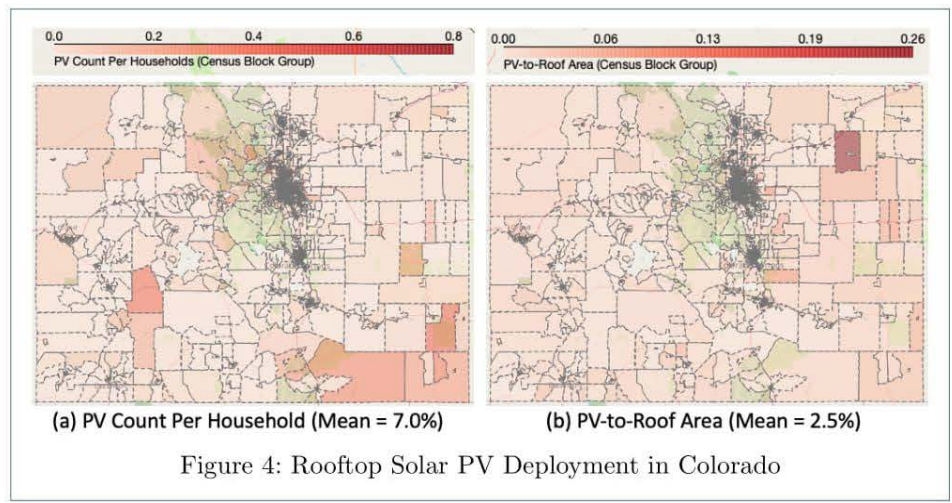

Figure 4: Rooftop Solar PV Deployment in Colorado

Table 1 reports the predictive performance of 16 ML models. Of the models predicting PV count per household, Model 5, including energy policy factors and trained on XGBoost, achieves the highest $R^2$ of 68.5%. Of the models predicting the PV-to-roof area ratio, Model 13, including energy policy factors and trained on XGBoost, has the best performance in terms of the $R^2$. XGBoost model with features that include energy policy factors explains 71.1% of the variance in the PV-to-roof area ratio.

Table 1: Model Performance Comparison

| Model | Dataset | Algorithm | MAE | RMSE | $R^2$ |
|---|---|---|---|---|---|
| M1 | PV count per HH | XGBoost | .038 | .003 | 62.2% |
| M2 | | CATBoost | .039 | .003 | 57.4% |
| M3 | | LightGBM | .038 | .003 | 60.4% |
| M4 | | Random Forest | .039 | 0.003 | 60.9% |
| M5 | PV count per HH + Energy policy | XGBoost | .038 | .003 | **68.5%** |
| M6 | | CATBoost | .008 | .0001 | 66.0% |
| M7 | | LightGBM | .008 | .0001 | 66.8% |
| M8 | | Random Forest | .009 | .001 | 61.8% |
| M9 | PV-to-roof ratio | XGBoost | .009 | .0002 | 55.7% |
| M10 | | CATBoost | .009 | .0002 | 56.0% |
| M11 | | LightGBM | .009 | .0001 | 59.2% |
| M12 | | Random Forest | .009 | .0002 | 56.0% |
| M13 | PV-to-roof ratio + Energy policy | XGBoost | .008 | .0001 | **71.1%** |
| M14 | | CATBoost | .008 | .0001 | 66.0% |
| M15 | | LightGBM | .008 | .0001 | 66.0% |
| M16 | | Random Forest | .009 | 0.0001 | 61.8% |

### 4.2 Predicting PV Count per Household

Figure 5 shows the ranking of the aggregated standardized feature importance scores (FIS) from the eight models predicting solar PV count per household. The FIS computed within each model was standardized to take a value between 0 and 1, and the $R^2$ of each model was used as a weight for aggregation. The bivariate correlation coefficient between each predictor and the target variable is annotated with colors indicating statistical significance.

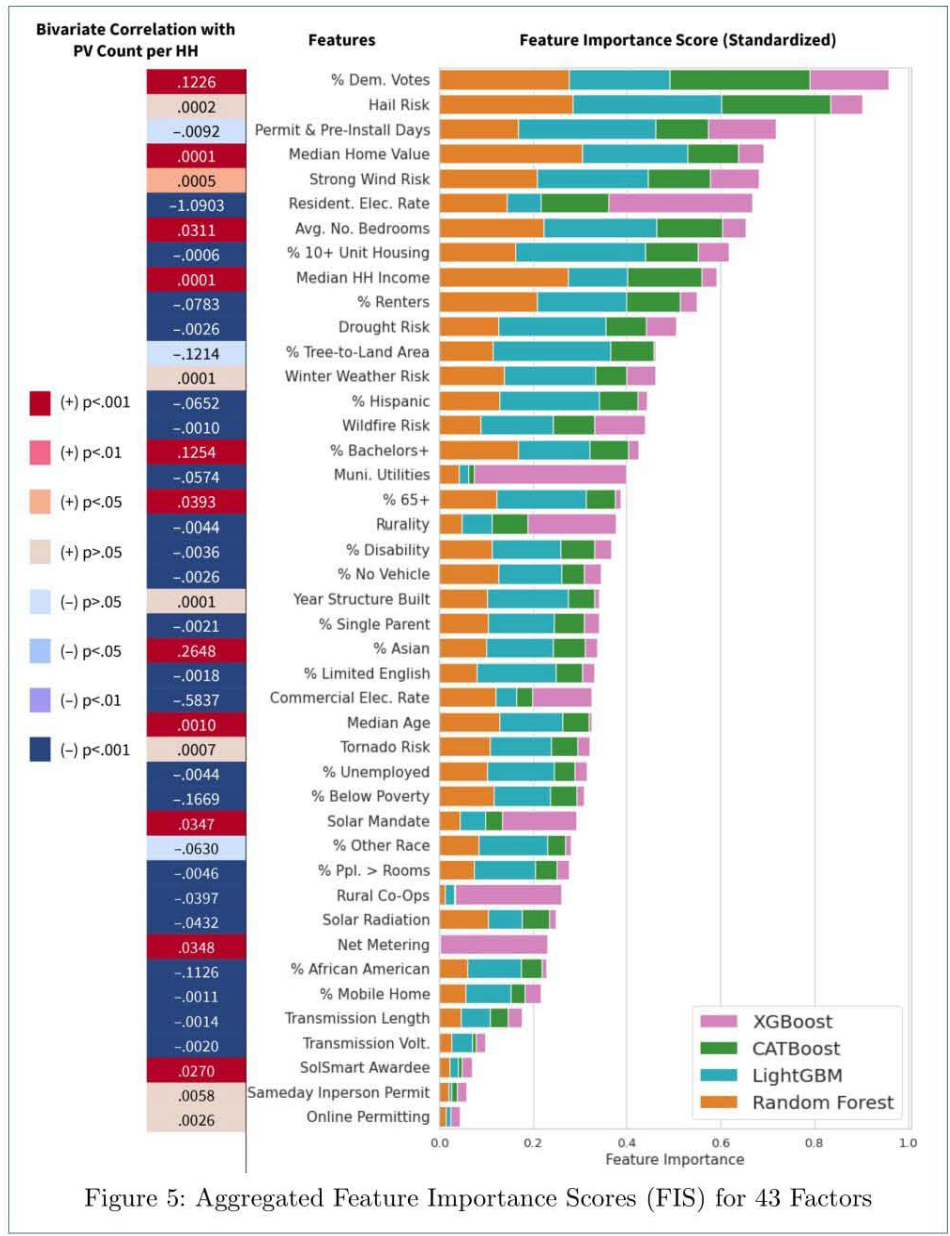

Figure 5: Aggregated Feature Importance Scores (FIS) for 43 Factors

Based on the aggregated FIS, the percent of Democratic presidential voters (% Dem. Votes) is the most important predictor of the PV count per household. The block groups with more Democrats have a greater number of PVs per household. The second most important predictor is hail risks. Block groups with a higher hail risk have a fewer PVs per household. The timeline for getting solar PV permits and pre-installation is the third most important predictor. The PV count per household is smaller in block groups with a longer expected duration between first permit submission and pre-installation completion.

Figure 6 presents the SHapley Additive exPlanations (SHAP) values, a unified measure developed to interpret the importance of input features [34], for 20 features most important in predicting PV count per household from XGBoost and LightGBM models, respectively. A positive SHAP value means that the feature has a positive impact on the target value, while a negative SHAP value means the opposite. The colors represent the relative contribution of that particular data point, or the observation, in predicting the outcome. Red pixels represent the higher importance of the data points in terms of the impact of the feature, while blue pixels represent the opposite.

Combining the results from the aggregated FIS (Figure 5) and the SHAP values (Figure 6), we explain how input features lead to spatial disparities in solar deployment below.

*4.2.1 Natural Environment*
The proportion of tree-to-land area and solar radiation are ranked 12th and 35th in the aggregated FIS scheme, respectively. PV count per household is greater in block groups with a higher tree-to-land area ratio. Contrary to expectations, solar radiation is slightly negatively associated with PV count per household in Colorado. However, this finding should not be generalized beyond Colorado. Most areas in the state have high solar potential (annual average DNI 4.5–7.5, which is 2–3 times higher than the world average DNI) and over 300 days of sun a year to ensure a highly functioning solar systems.

*4.2.2 Demographics and Built Environments*
Median household income positively predicts PV count per household. Although median household income is not the most important predictor of all 43 factors, other income-related variables, such as median home value and the number of bedrooms, are ranked high in the aggregated FIS scheme and have higher SHAP values. In all models, median home value and the number of bedrooms are found to have positive impact on PV count per household.

Racial and ethnic disparities in solar deployment exist across block groups. The share of Asians positively predicts PV count per household, while the share of Hispanics negatively predicts PV count per household. Our supplementary statistical analysis report similar findings (Appendix Table C2). Block groups with a greater share of African Americans have a fewer PV count per household, even after controlling for median household income, home value, and house sizes.

The results from the supplementary analysis Model 20 (Appendix Table C2) with interaction terms between race and median household income suggest differential marginal effects of median household income on PV count per household across neighborhoods with different racial compositions. Figure 7 shows the increase of Asian and White populations is associated with the positive average marginal effects of income on PV count per household, while the increase of Black and Hispanic populations suggests otherwise.

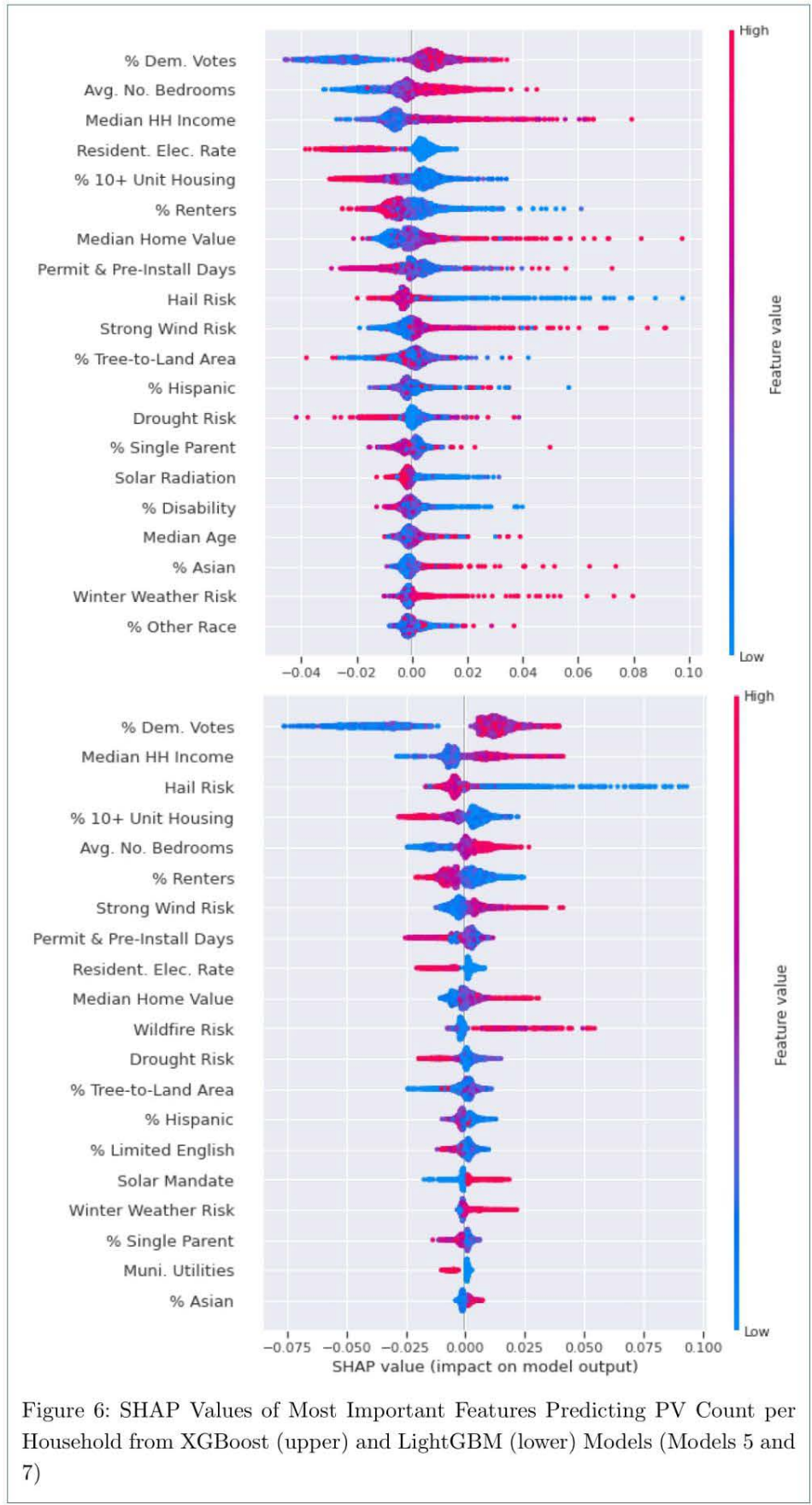

Figure 6: SHAP Values of Most Important Features Predicting PV Count per Household from XGBoost (upper) and LightGBM (lower) Models (Models 5 and 7)

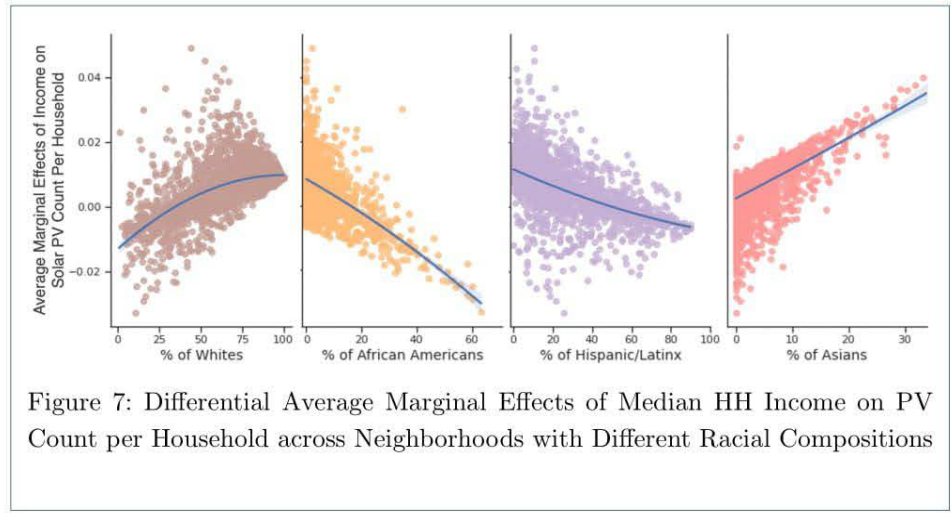

Figure 7: Differential Average Marginal Effects of Median HH Income on PV Count per Household across Neighborhoods with Different Racial Compositions

The results provide clear evidence that political views impact PV count per household. The share of Democratic party votes positively predicts PV count per household across all models and is the first- or second-most important predictor across Models 1–8.

4.2.3 *Energy Infrastructure, Market, and Policy*
Although the lengths and aggregated voltage of transmission lines in block groups negatively relate to PV count per household (Appendix Table C3), these transmission factors are not as important as other variables based on the results from the ML models and SHAP analysis.

The results from the SHAP analysis indicate that PV count per household is lower in areas with higher electricity rates, suggesting energy cost savings may not be the main driver of consumer-level rooftop solar system adoption in Colorado. But this finding should not be generalized because the variation of electricity price across block groups in Colorado is lower than that in the United States.

Overall, neighborhoods in municipal utility service areas and rural co-ops have lower PV count per household compared to investor-owned utility service areas. But when energy policy variables are included, this difference across utility service areas becomes insignificant, suggesting that the disparities in PV deployment are driven more by utility policies, than by ownership type.

Of the energy policy variables, permit and pre-install days and solar mandate appear to be the two most important predictors of rooftop solar adoption. The median business days between first permit submission to the local jurisdiction and pre-install interconnection negatively predict PV count per household, indicating that solar adoption may be discouraged in areas with longer timelines for solar permit approval. The results provide clear evidence supporting neighborhoods in cities or counties with solar mandates have a greater number of per-household PVs. Net metering has a statistically significant and positive effect on solar deployment, but the relative importance of net metering is very low as most utilities in Colorado offer net metering programs.

*4.2.4 Social and Natural Disaster Vulnerabilities*

PV count per household is lower in block groups with a greater share of single parents and persons with disabilities, even after controlling for household income and median home value. Other social vulnerability factors, such as the proportion of households below the poverty level and without vehicles, that are highly associated with median household income and home values, appear less important than other social vulnerability factors.

Natural disaster vulnerability is one of the most important predictors of PV adoption, but the directions of the effects vary across the types of natural disasters. Based on the results from both ML and statistical models (Appendix Table C4), strong wind and winter weather risks positively predict PV count per household, suggesting neighborhoods with higher power outage frequencies due to wind and snowstorms may be more likely to deploy PV systems, compared to their counterparts. However, block groups with higher hail risks are less likely to deploy PVs, indicating concerns over hail damages to the solar panels could be an actual barrier to rooftop PV adoption. The direction of drought and wildfire risks are inconclusive, but both stay ranked higher than the median in comparison.

4.3 Predicting PV-to-Roof Area

Figure 8 displays the ranking of aggregated standardized FIS from the eight models predicting PV-to-roof area in each block group. The standardized FIS is weighted by the explained variance ($R^2$) of each model. The bivariate correlation coefficient is annotated with different colors indicating statistical significance.

The average number of bedrooms, which is a proxy for the average size of houses, is the most important predictor of the PV-to-roof area ratio. Not surprisingly, the average number of bedrooms negatively predicts the PV-to-roof area ratio because the size of the roof area is highly correlated with the average number of bedrooms. The second most important predictor is the permit and pre-install duration, meaning the PV-to-roof area ratio is smaller in block groups which have longer timelines for getting the permits and pre-installation of PV projects. The third most important predictor is the percentage of renters. A high percentage of renters is associated with a higher PV-to-roof area ratio. This result also makes sense as neighborhoods with more rental units generally have smaller roof areas.

Figure 9 reports SHAP values for 20 of the most important features in predicting PV-to-roof area from XGBoost and LightGBM models, respectively. The relative importance and the impact of each predictor are explained below.

*4.3.1 Natural Environment*

Overall, the findings for the natural environment features are similar to the findings from the analysis of the PV count per household. The proportion of tree-to-land area is negatively associated with the PV-to-roof area ratio, suggesting that block groups with more tree coverage have smaller PVs relative to the size of the roofs. This negative relationship between the tree-to-land area and PV-to-roof area ratios is statistically significant, even after controlling for all other determinants (Model 34 in Appendix Table C5). Solar radiation negatively predicts the ratio of PV-to-roof area. But this finding should not be generalized beyond Colorado because the state is an exception with abundant solar radiation to ensure highly effective energy generation in most places in the state.

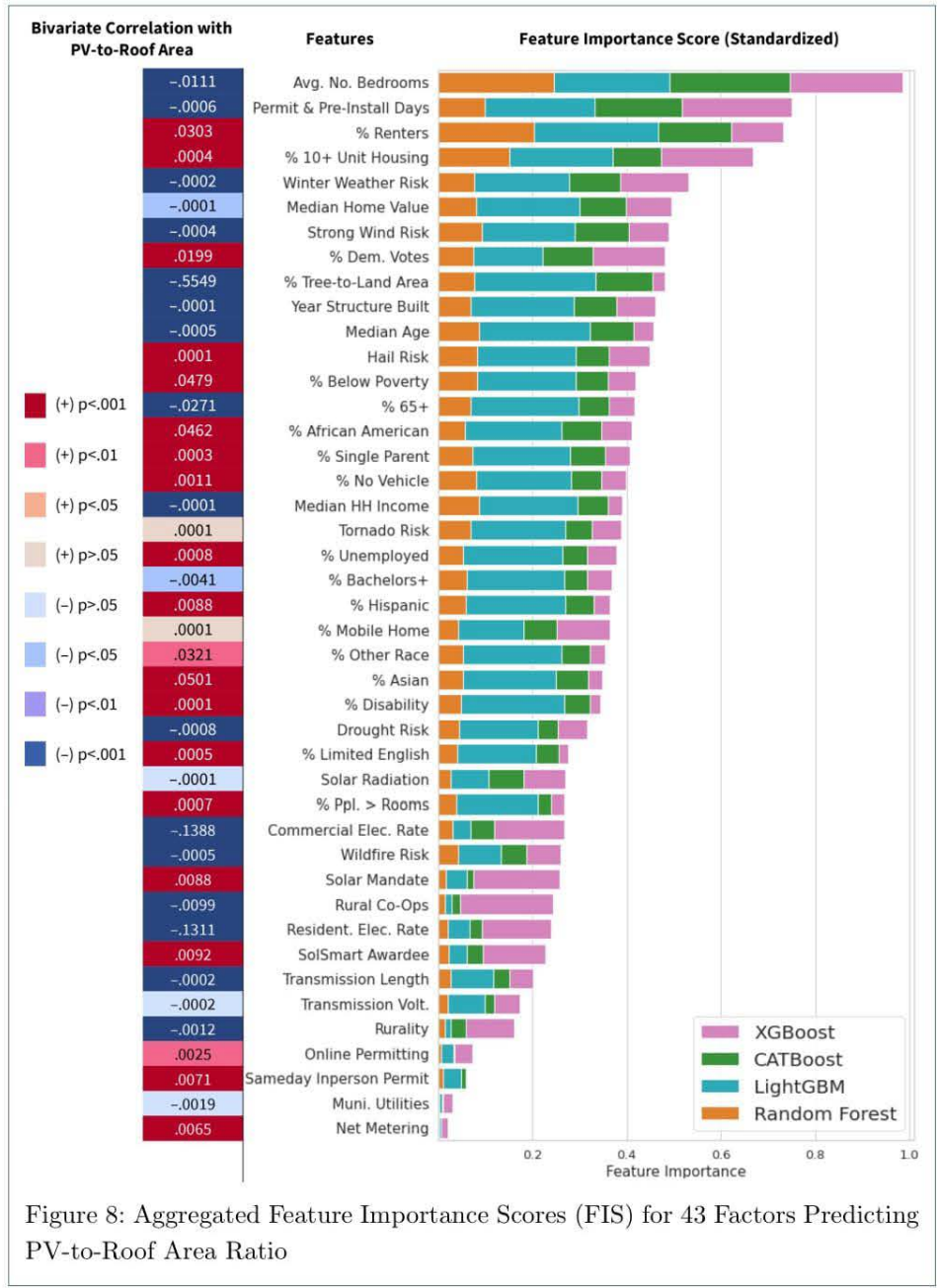

Figure 8: Aggregated Feature Importance Scores (FIS) for 43 Factors Predicting PV-to-Roof Area Ratio

#### 4.3.2 Demographics and Built Environment

The findings for demographics and built environment should be interpreted with caution as many demographic and built environment predictors can be associated with both PV and roof areas. If this is the case, the direction of SHAP values for a predictor will be determined by which of the two –PV or roof areas– is *more* impacted by the predictor. For example, even if the median home value positively predicts PV areas in a neighborhood, PV-to-roof ratio can be negatively associated with median home value because roof areas are significantly larger in areas with a higher median home value.

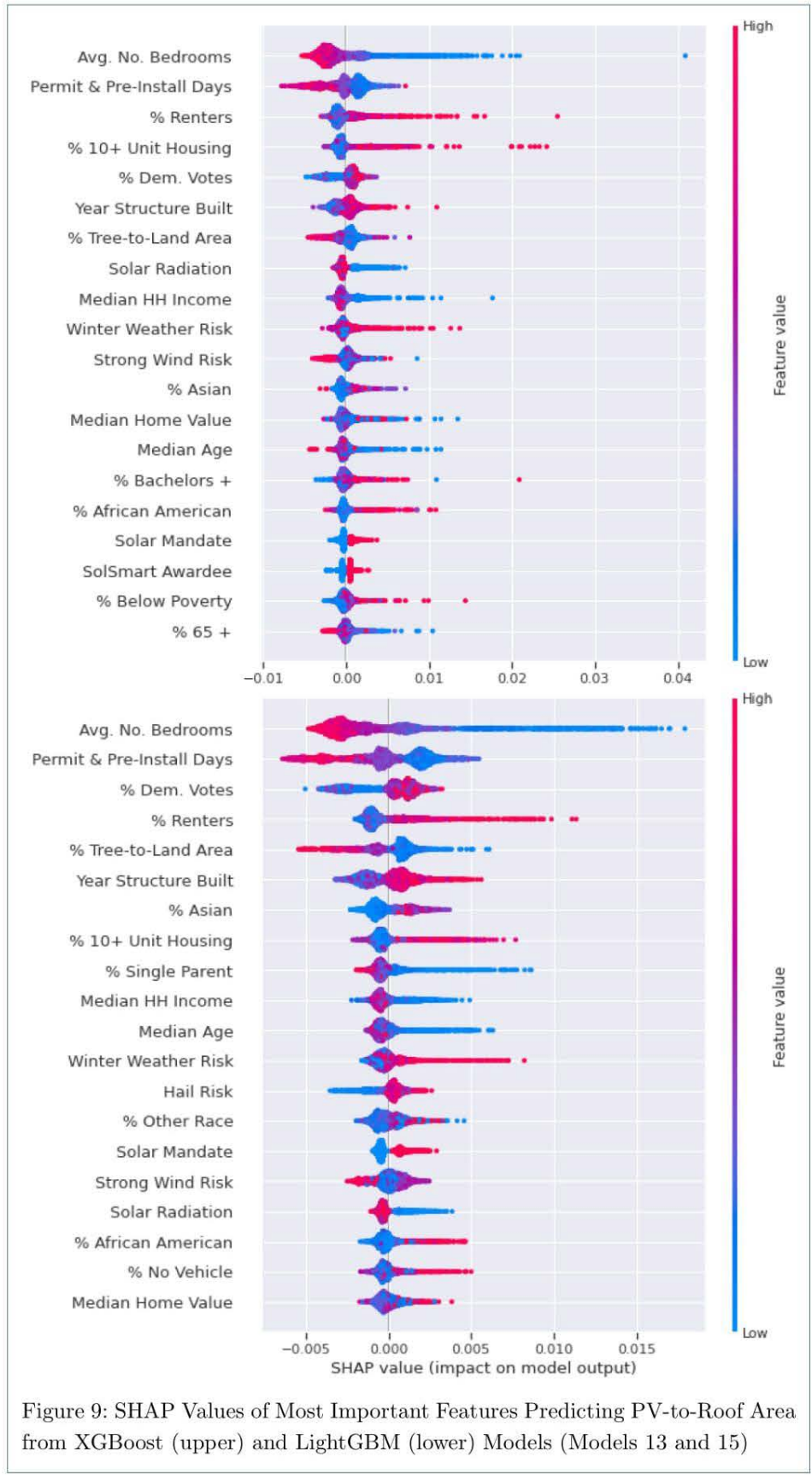

Figure 9: SHAP Values of Most Important Features Predicting PV-to-Roof Area from XGBoost (upper) and LightGBM (lower) Models (Models 13 and 15)

The proportion of renters and multi-dwelling unit houses are the third and fourth most important and positive predictors of the PV-to-roof area ratio, suggesting that neighborhoods with more renters and multi-dwelling units have proportionally more roof areas covered by PVs. Median household income and home values are negatively associated with PV-to-roof areas, suggesting a positive association between income or home values and roof sizes may outweigh the negative association between those two. Consistent with the findings from the PV-count-per-household analysis above, block groups with more Democrats have a higher PV-to-roof ratio. Based on the results from both ML and supplementary statistical analysis (Appendix Table C2), median age is found to have a negative impact on the PV-to-roof ratio.

Racial disparities in the PV-to-roof area ratio are explained by the proportion of Blacks and Asians. Based on the SHAP values from the XGBoost model, the distribution of the PV-to-roof area ratio approximates to the bimodel distribution between block groups with high and low Black populations. Neighborhoods with high or low share of Black residents have a higher PV-to-roof area ratio, while neighborhoods with moderate share of Black residents have a lower ratio. Based on both SHAP and statistical analyses, block groups with a higher share of Asians tend to have a greater PV-to-roof area ratio. Unlike the results from the previous analysis on PV count per household, the interaction between income and racial composition is less salient.

*4.3.3 Energy Infrastructure, Market, and Policy*

The average permit and pre-install timelines are the second most important and negative predictor of the PV-to-roof area ratio. This suggests that generally speaking, PV permitting and inspection timelines can affect not only the decision to adopt rooftop solar but also the size of the deployed PV systems. According to the SHAP analysis, the feature importance of solar mandates is not as high as other predictors, but the positive impact of solar mandates on the PV-to-roof area is still clear. Other energy infrastructure, market, and policy variables (e.g., net metering, online permitting, transmission voltage and length, and utility ownership) are ranked lower in the FIS spectrum. Relatively speaking, energy infrastructure, market, and policies appear to have a more notable impact on PV system adoption than the PV system size.

*4.3.4 Social and Natural Disaster Vulnerabilities*

We do not find any significant evidence to suggest that social vulnerabilities impact the PV sizes. Only consistent, and obvious findings are that the PV-to-roof area ratio is higher in block groups with a greater share of households below poverty and that of households who live in 10+ unit housing. This could be simply because these two features are highly correlated with the size of the house. Based on the ML models and SHAP analysis, block groups with a higher proportion of single-parent household have a lower PV-to-roof ratio, but this finding is not supported by the supplementary statistical analysis. The findings for other social vulnerability variables (i.e., % disability, % limited English, % no vehicle, % number of people > number of rooms, % no vehicles, % unemployed) are inconclusive based on the aggregated FIS, SHAP analysis (Appendix Table C4), and supplementary statistical analysis.

Of the natural disaster vulnerability features, winter weather risk and wind risk are the important features predicting the PV-to-roof ratio. Block groups with higher winter weather risk have a higher PV-to-roof ratio, suggesting residents in areas with frequent snow storms are more likely to deploy larger PV systems relative to the size of the roofs. The second most important natural disaster vulnerability feature is strong wind risk. Block groups with a high strong wind risk have a lower PV-to-roof ratio, which is contrary to the findings for the PV count per household. These findings on winter weather and strong wind risks are also supported by the supplementary analysis in Appendix Table C4. Findings for other natural disaster vulnerability factors including drought, wildfire, hail, and tornado risks, remain inconclusive.

## 5 Discussion and Conclusion

Advances in the real-time data flow provide ample opportunities for innovative research on addressing important social issues, such as sustainable and healthy cities [5, 14, 31], energy efficiency [8], emergency management [25], and social justice [52]. Major progress in the availability and quality of satellite imagery has opened up the possibility to detect the spatial distribution of distributed energy resources (DERs), such as solar PV, wind generating units, and outdoor battery storage, in near real time. This study demonstrates the value of data science in discovering the patterns and trends in DER deployment and providing useful insights for guiding decision making around energy infrastructure investments that can support sustainable, resilient, and equitable energy systems.

Although previous work has shown that spatial disparities in solar PV deployment exist [33, 48], we lacked a comprehensive understanding of how natural and built environment, disaster vulnerabilities, and PII rules, along with socioeconomic factors, jointly create such disparities. Without this understanding, the mechanisms behind the persisting socioeconomic gap in solar remains unclear. Our investigation of rooftop solar in Colorado provides a more complete picture of how natural and social features are deeply interwoven and lead to spatial disparities in PV deployment and suggests a way to create a more accurate PV deployment prediction model.

Combining the results from the ML models, SHAP analyses, and statistical estimations, this study offers two major implications on energy policy and future DER investment in consideration of energy justice and social equity. First, the results of this study suggest that while strategic adoption of solar policies such as PII rules and solar mandates can increase the number of per-household PVs, significant disparities continue to exist among adopters and non-adopters and in turn those who are benefiting and not benefiting from solar energy. State and local governments' energy regulations are frequently codified as alternative energy policy with rooftop solar advocated as a method for increasing the use of DERs and overall grid reliability while simultaneously resulting in lower utility bills and associated health and social benefits. Aligning with previous research [41, 48], however, this study finds that individuals who have limited English proficiency, who are African American or Hispanic, and who are lower-income deploy solar at a lower rate than their English-speaking, Asian, White, and higher-income counterparts. Solar disparities have been attributed to multiple interrelated factors such as homeownership rates, internet access, information and knowledge gaps, and the ability to participate in environmental decision making. Given that low-income households, people of color, and renters are disproportionately more susceptible to energy vulnerability [23, 28], our findings suggest that state and local governments should consider the underlying social and demographic conditions that may inhibit rooftop solar adoption to ensure equitable grid infrastructure investment for DERs.

Second, as cities increasingly adopt new building codes and permitting rules to promote rooftop solar systems, local governments need to be cognizant of who benefits more and who benefits less from the transition to renewable energy. Current solar policies overwhelmingly benefit groups that can afford solar, own their own homes, and are middle- to upper-income. For example, solar mandates have the potential to reduce the installation costs of PV systems through consideration of solar in the initial design of a building or through installation of solar at the time of construction; however, low-income and people of color are significantly less likely to own their own homes. In areas with a higher portion of renters and multi-family dwelling unit residents, incentivizing the deployment of rooftop solar for rental properties may address the problems associated with split-incentives in which landlords assume the costs of energy efficient property upgrades without the reduced utility benefits experienced by renters [10, 35]. Since current financial incentives such as the Federal Solar Tax Credit do not apply to rental properties, landlords have little motivation to deploy rooftop solar even when their roofs have the technical capacity to do so. Additionally, increased development and access to affordable community solar farms can allow those who have limited resources or are unable to install rooftop solar the ability to subscribe to a shared PV system and as such experience the same benefits as those who deploy rooftop solar.

Besides the implications for energy justice and social equity, our study sheds light on the importance of energy policies for promoting rooftop solar in general. Across all models, our results suggest that reducing permit and pre-installation timelines and solar mandates enhance rooftop solar deployment significantly, while online and same-day in-person permits are relatively less important. The soft costs, or non-hardware costs including PII, financing, and knowledge acquisition costs, can be critical as they account for the 50 to 70 percent of total installed PV system price [6, 37]. City and county governments that wish to promote rooftop solar deployment should consider lowering barriers for solar permitting and interconnection, especially to reduce the time availability and information gap that may exist across social groups under varying geospatial and socioeconomic circumstances.

Our study also offers implications for climate- and environment-specific strategies for DER development. Across our results, we find that tree canopy cover, hail risks, winter weather risks, strong wind risks, and tornado risks are important predictors of solar PV deployment in Colorado. Although these natural conditions are not easy to modify, they should be taken into consideration in grid infrastructure investment for enhancing community resilience and incorporating different renewable resources into the energy portfolio. For example, areas with a high risk of power outages due to heavy snow may be more interested in solar systems, and large-scale investment in batteries can be beneficial for enhancing power grid resilience in those areas. Areas with high hail risks or tornado risks may be less desirable for rooftop solar, and other types of DERs could be considered as an alternative. With solar panels becoming more resilient to hail storms and wind, however, educating potential PV adopters about the durability and safety of solar panels could be more important in promoting PV deployment in those areas.

We note that our data are limited to Colorado, USA, and are only from 2021. The findings from our analyses may not be generalizable to other states or countries. In addition, to investigate the effects of the input features on solar deployment over time, multiple years of satellite imagery data showing the different patterns of PV deployment diffusion would be necessary. Future studies should seek to measure block- or block group-level spatial disparities in solar deployment over multiple years to fully uncover the underlying dynamics of the spatial distribution of solar PV deployment.

We encountered several challenges while training with Faster-RCNN models [16, 49]. The RPN layer has exhibited poor performance for small sized object localization due to the coarse grained nature of satellite images. We find similar behavior in our model as illustrated in figure D, an example run of training to identify roofs. Additionally, the RPN layer has difficulty in distinguishing PVs from background effects such as shadows and contrast changes of roofs due to less variation in color leading to some false detection. Moreover, the IoU threshold used could also potentially affect the performance of the model as discussed in [24]. The available manually annotated data for satellite detection is relatively small and could potentially lead to over-fitting issues. Finally, our model outputs rectangular frames surrounding the objects of interest and we are interested in calculating the ratio of the area of the object of interest to that of the whole image, this introduces another source of error since the object of interest does not necessarily occupy the predicted frame in entirety. Some studies that have tried to address issues related to dense, small and arbitrary rotations of objects can be found in [22, 54]. Future studies may be able to enhance the performance of computer vision models for detecting roof and PV areas by addressing some of these challenges.

Despite these limitations, our study is useful because this study illustrates how to effectively extract meaningful solar PV deployment data from the massive publicly available satellite imagery data sources. This study also generates a novel and granular solar deployment dataset. Although this new dataset is specific to Colorado, USA, our approach demonstrates how to produce a comprehensive and representative dataset of solar PV deployment in other parts of the world. Moreover, we developed one of the most accurate ML models on solar PV deployment to date, predicting over 70% of the variation in the deployment. This study also provides new insights on how spatial distribution of solar deployment depends on natural disasters, tree canopy, social vulnerabilities, transmission infrastructure, and permitting and interconnection processes, among others.

Kim *et al.*　　　Page 22 of 29## APPENDIX A. Summary Statistics

**Table A1** Dataset Summary Statistics

|                         | Count | Mean    | STD    | Min    | Max     |
|-------------------------|-------|---------|--------|--------|---------|
| PV Count per HH         | 3441  | .070    | .093   | .001   | .784    |
| PV-to-Roof Ratio        | 3441  | .025    | .020   | .001   | .259    |
| % Tree-to-Land Area     | 3441  | .006    | .007   | .001   | .052    |
| Solar Radiation         | 3441  | 5.851   | .430   | 4.712  | 7.236   |
| Median HH Income        | 3441  | 77571   | 36263  | 14145  | 250000  |
| Median Age              | 3441  | 39.608  | 9.067  | 17.7   | 84.6    |
| % 65 +                  | 3441  | .268    | .133   | .001   | 1.000   |
| % Bachelors +           | 3441  | .283    | .164   | .001   | .855    |
| % Renters               | 3441  | .333    | .253   | .001   | 1.000   |
| Year Structure Built    | 3441  | 1977    | 17     | 1939   | 2014    |
| Avg. No. Bedrooms       | 3441  | 2.872   | .664   | .466   | 4.524   |
| Median Home Value       | 3441  | 354098  | 200610 | 10000  | 2000000 |
| Rurality                | 3441  | 2.098   | 2.429  | 1.000  | 9.000   |
| % Dem. Votes            | 3441  | .553    | .154   | .109   | .796    |
| % African American      | 3441  | .035    | .073   | .000   | .633    |
| % Hispanic              | 3441  | .213    | .197   | .000   | .923    |
| % Asian                 | 3441  | .027    | .043   | .000   | .422    |
| % Other Race            | 3441  | .032    | .042   | .000   | .973    |
| Transmission Volt.      | 3441  | 1.952   | 2.724  | .000   | 8.166   |
| Transmission Length     | 3441  | 4.739   | 5.791  | .000   | 14.455  |
| Muni. Utilities         | 3441  | .096    | .295   | .000   | 1.000   |
| Rural Co-Ops            | 3441  | .215    | .411   | .000   | 1.000   |
| Resident. Elec. Rate    | 3441  | .122    | .019   | .062   | .212    |
| Commercial Elec. Rate   | 3441  | .103    | .017   | .076   | .260    |
| Solar Mandate           | 3076  | .271    | .444   | .000   | 1.000   |
| Net Metering            | 3441  | .919    | .273   | .000   | 1.000   |
| SolSmart Awardee        | 2339  | .501    | .500   | .000   | 1.000   |
| Online Permit           | 2339  | .754    | .431   | .000   | 1.000   |
| Sameday InPerson Permit | 2339  | .435    | .496   | .000   | 1.000   |
| Permit & Pre-Install Days | 2339 | 16.267 | 6.002  | 8.000  | 35      |
| Drought Risk            | 3441  | 1.756   | 3.285  | .000   | 29.35   |
| Wildfire Risk           | 3441  | 4.712   | 8.509  | .000   | 48.921  |
| Hail Risk               | 3441  | 28.292  | 13.027 | 2.583  | 64.351  |
| Winter Weather Risk     | 3441  | 13.037  | 9.868  | .000   | 62.890  |
| Strong Wind Risk        | 3441  | 17.077  | 7.332  | 4.107  | 59.476  |
| Tornado Risk            | 3441  | 30.703  | 9.831  | 5.345  | 56.195  |
| % Below Poverty         | 3441  | .105    | .099   | .000   | .847    |
| % Disability            | 3441  | 11.281  | 5.050  | .400   | 44.6    |
| % Single Parent         | 3441  | 7.884   | 4.738  | .000   | 27.600  |
| % Limited English       | 3441  | 3.028   | 4.317  | .000   | 37.700  |
| % 10+ Unit Housing      | 3441  | 14.635  | 18.589 | .000   | 98.900  |
| % Mobile Homes          | 3441  | 4.329   | 8.638  | .000   | 79.100  |
| % Ppl. > Rooms          | 3441  | 2.767   | 3.303  | .000   | 24.800  |
| % No Vehicle            | 3441  | 5.217   | 5.310  | .000   | 43.300  |
| % Unemployed            | 3441  | 4.861   | 2.995  | .000   | 28.400  |

# APPENDIX B. Hyperparameters Used in Predictive Models

**Table B1** Hyperparameters used in each ML algorithm

| Model | Dataset | Algorithm | Hyperparameters |
|---|---|---|---|
| M1 | | XGBoost | 'gamma': 0, 'alpha': 12, 'learning_rate': .027, 'seed': 712 'colsample_bytree': .3, 'reg_lambda': 1,'random_state': 700, 'n_estimators': 299, 'base_score': .29, 'max_depth': 7 |
| M2 | | CATBoost | 'l2_leaf_reg': 2, 'learning_rate': 0.1, 'depth': 9, 'iterations': 150 |
| M3 | PV Count per HH | LightGBM | 'objective': 'regression', 'metric': 'rmse','is_unbalance': 'true', 'is_training_metric': 'true', 'boosting': 'gbdt', 'num_leaves': 36, 'feature_fraction': .99, 'bagging_fraction': .69, 'bagging_freq': 4, 'learning_rate': .01, 'max_depth': 15, 'max_bin': 23 |
| M4 | | RandomForest | 'n_estimators': 19, 'max_depth': 150, 'min_samples_split': 2, 'max_features': "sqrt",'min_samples_leaf': 2, 'random_state': 531 |
| M5 | | XGBoost | 'gamma': 0, 'alpha': 5, 'learning_rate': .05, 'random_state': 185, 'colsample_bytree': .5, 'reg_lambda': 0, 'n_estimators': 311, 'base_score': .5, 'max_depth': 7, 'seed': 855 |
| M6 | | CATBoost | 'l2_leaf_reg': 2, 'learning_rate': 0.1, 'depth': 6, 'iterations': 200 |
| M7 | PV Count per HH +Energy Policy | LightGBM | 'objective': 'regression', 'metric': 'rmse','is_unbalance': 'true', 'is_training_metric': 'true', 'boosting': 'gbdt', 'num_leaves': 36, 'feature_fraction': .81, 'bagging_fraction': .91, 'bagging_freq': 20, 'learning_rate': .021, 'max_depth': 14, 'max_bin': 23 |
| M8 | | RandomForest | 'n_estimators': 700, 'max_depth': 150, 'min_samples_split': 2, 'max_features': "sqrt",'min_samples_leaf': 2, 'random_state': 372 |
| M9 | | XGBoost | 'gamma': 0, 'alpha': 12, 'learning_rate': .025, 'seed':712 'colsample_bytree': .35, 'reg_lambda': 1, 'random_state': 789, 'n_estimators':300, 'base_score': .5, 'max_depth': 8 |
| M10 | | CATBoost | 'l2_leaf_reg': 1, 'learning_rate': 0.09, 'depth': 10, 'iterations': 200 |
| M11 | PV-to-Roof Area | LightGBM | 'objective': 'regression', 'metric': 'rmse','is_unbalance': 'true', 'is_training_metric': 'true', 'boosting': 'gbdt', 'num_leaves': 45, 'feature_fraction': .25, 'bagging_fraction': .75, 'bagging_freq': 4, 'learning_rate': .01, 'max_depth': 15, 'max_bin': 52 |
| M12 | | RandomForest | 'n_estimators': 300, 'max_depth': 64, 'min_samples_split': 3, 'max_features': sqrt, 'min_samples_leaf': 2, 'random_state': 435 |
| M13 | | XGBoost | 'gamma': 0, 'alpha': 5, 'learning_rate': .05, 'seed': 1164 'colsample_bytree': .5, 'reg_lambda': 0,'random_state': 185, 'n_estimators': 500, 'base_score': .52, 'max_depth': 9 |
| M14 | | CATBoost | 'l2_leaf_reg': 1, 'learning_rate': 0.09, 'depth': 6, 'iterations': 150 |
| M15 | PV-to-Roof Area +Energy Policy | LightGBM | 'objective': 'regression', 'metric': 'rmse','is_unbalance': 'true', 'is_training_metric': 'true', 'boosting': 'gbdt', 'num_leaves': 36, 'feature_fraction': .34, 'bagging_fraction': .75, 'bagging_freq': 4, 'learning_rate': .01, 'max_depth': 15, 'max_bin': 23 |
| M16 | | RandomForest | 'n_estimators': 300, 'max_depth': 280, 'min_samples_split': 2, 'max_features': sqrt, 'min_samples_leaf': 2, 'random_state': 42 |

# APPENDIX C. Sub-Analysis Linear Regression Models

**Table C1** Tree Canopy, Solar Radiation, and PV-to-Roof Area

| Predictors | PV Count per HH Model 17 Coef. (SE) | PV-to-Roof Area Model 18 Coef. (SE) |
|---|---|---|
| % Tree-to-Land Area | 2.636 (.523)*** | −.537(.111)*** |
| % Tree-to-Land Area$^2$ | -116 (17.59)*** | −1.111(3.744) |
| Solar Radiation (log) | −.053 (.004)*** | −.001(.001)* |
| Solar Radiation (log)$^2$ | -.106 (.008)*** | −.004(.002)* |
| Number of Block Groups | 3441 | 3441 |
| $R^2$ | .052 | .041 |

*** $p < 0.001$. Robust standard errors (SE) are in parentheses.

**Table C2** Demographics, Built Environments, and Solar PV Deployment

| | PV Count per Household | | PV-to-Roof Area | |
|---|---|---|---|---|
| Predictors | Model 19 Coef. (SE) | Model 20 Coef. (SE) | Model 21 Coef. (SE) | Model 22 Coef. (SE) |
| Median HH Income | .001 (.001) | .001 (.001) | −.001 (.001)*** | −.001 (.001) |
| Median HH Income$^2$ | −.001 (.001) | | .001 (.001)*** | |
| Median Age | −.001 (.001) | −.001 (.001) | −.001 (.001)*** | −.001 (.001)*** |
| % 65+ | .010 (.018) | .009 (.018) | .004 (.004) | .005 (.004) |
| % Bachelor+ | −.014 (.018) | −.015 (.018) | .004 (.004) | .001 (.004) |
| % Renters | −.049 (.012)*** | .049 (.012)*** | .004 (.003)† | .006 (.003)* |
| Year Structure Built | .100 (.017)*** | .105 (.017)*** | .011 (.004)*** | .011 (.004)*** |
| Year Structure Built$^2$ | −.001 (.001)*** | −.001 (.001)*** | −.001 (.001)*** | −.001 (.001)*** |
| Avg. No. Bedrooms | .011 (.005)* | .011 (.005)* | −.009 (.001)*** | −.009 (.001)*** |
| Median Home Value | .001 (.001)*** | .001 (.001)*** | .001 (.001)** | .001 (.001)** |
| Rurality | −.001 (.001)† | −.001 (.001)† | −.001 (.001)*** | −.001 (.001)*** |
| % Dem. Voters | .107 (.014)*** | .115 (.014)*** | .002 (.003) | .001 (.003) |
| % African American | −.101 (.022)*** | .420 (.561) | .014 (.005)** | .083 (.115) |
| % Hispanic | −.026 (.011)* | .409 (.179)* | −.002 (.002) | .073 (.037)* |
| % Asian | .220 (.037)*** | −3.245 (.818)*** | .021 (.008)** | .036 (.168) |
| % Other Race | .008 (.037) | −.575 (.657) | .010 (.008) | .001 (.135) |
| Income × Afri. Ameri. | | −.048 (.051) | | −.006 (.010) |
| Income × Hispanic | | −.041 (.017)* | | −.007 (.003)* |
| Income × Asian | | .310 (.073)*** | | −.001 (.015) |
| Income × Other Race | | .053 (.061) | | .001 (.013) |
| No. of block groups | 3441 | 3441 | 3441 | 3441 |
| $R^2$ | .143 | .149 | .202 | .198 |

*** $p < 0.001$, ** $p < 0.01$, * $p < 0.05$, † $p < 0.1$. Robust Standard Errors are in parentheses.

Table C3 Impact of Energy Market, Policy, and Infrastructure on Solar PV Deployment

|  | Solar PV Count per Household | | PV-to-Roof Area | |
| --- | --- | --- | --- | --- |
|  | Model 23 | Model 24 | Model 25 | Model 26 |
| Predictors | Coef. (SE) | Coef. (SE) | Coef. (SE) | Coef. (SE) |
| Transmission Volt. | −.003 (.001)** | .001 (.003) | .001 (.001) | −.001 (.001) |
| Transmission Length | .001 (.001) | −.001 (.001) | −.001 (.001) | .001 (.001) |
| Muni. Utilities | −.048 (.005)*** | −.068 (.009)*** | −.005 (.001)*** | .004 (.003) |
| Rural Co-ops. | −.046 (.005)*** | −.005 (.011) | −.008 (.001)*** | .002 (.004) |
| Resident. Elec. Rate | −.839 (.140)*** | .150 (.348) | .003 (.029) | .582 (.106)*** |
| Commerc. Elec. Rate | 1.067 (.165)*** | −2.06 (.648)** | .043 (.034) | −.123 (.139) |
| Solar Mandate |  | .001 (.009) |  | .006 (.002)** |
| Net Metering |  | .027 (.012)* |  | .008 (.008)** |
| SolSmart Awardee |  | .003 (.007) |  | −.001 (.001) |
| Online Permitting |  | .004 (.005) |  | −.001 (.001) |
| Sameday Inperson Per. |  | .002 (.007) |  | −.004 (.001) |
| Per. & Pre-Install Days |  | −.001 (.004)* |  | −.001 (.001)*** |
| CONTROLS | YES | YES | YES | YES |
| No. of block groups | 3441 | 2328 | 3441 | 2328 |
| $R^2$ | .216 | .277 | .234 | .325 |

*** $p < 0.001$, ** $p < 0.01$, * $p < 0.05$, † $p < 0.1$. Robust Standard Errors are in parentheses.
CONTROLS include Income, Median Age, % 65+, % Bachelor+, % Renters, Year Structure Built, Avg. No. Bedrooms, Median Home Value, Rurality, % Dem. Voters, % African American, % Hispanic, % Asian, % Other Race, % Tree-to-Land Area, Solar Radiation

Table C4 Social and Natural Disaster Vulnerabilities and Solar PV Deployment

|  | PV Count Per Household | | PV-to-Roof Area | |
| --- | --- | --- | --- | --- |
|  | Model 27 | Model 28 | Model 29 | Model 30 |
| Predictors | Coef. (SE) | Coef. (SE) | Coef. (SE) | Coef. (SE) |
| Drought Risk | .001 (.001) |  | .001 (.001) |  |
| Wildfire Risk | −.001 (.001)*** |  | −.001 (.001)* |  |
| Hail Risk | −.001 (.001)† |  | .001 (.001) |  |
| Winter Weather Risk | .001 (.001) |  | .001 (.001)*** |  |
| Strong Wind Risk | .002 (.001)*** |  | −.002 (.001)*** |  |
| Tornado Risk | −.002 (.001)*** |  | −.003 (.001)*** |  |
| % Below Poverty |  | .001 (.001) |  | .015 (.001)* |
| % Disability |  | −.001 (.001) |  | −.001 (.001) |
| % Single Parent |  | .001 (.001) |  | −.001 (.001) |
| % Limit. English |  | .001 (.001) |  | .001 (.001) |
| % 10+ Unit Housing |  | −.001 (.001)*** |  | .001 (.001)*** |
| % Mobile Home |  | .001 (.001) |  | .001 (.001)** |
| % Ppl. > Rooms |  | −.002 (.001)*** |  | −.001 (.001)*** |
| % No Vehicles |  | −.001 (.001) |  | .001 (.001) |
| % Unemployed |  | .001 (.001) |  | .001 (.001)*** |
| CONTROLS | YES | YES | YES | YES |
| No. of block groups | 3441 | 3441 | 3441 | 3441 |
| $R^2$ | .181 | .174 | .231 | .242 |

*** $p < 0.001$, ** $p < 0.01$, * $p < 0.05$, † $p < 0.1$. Robust Standard Errors are in parentheses.
CONTROLS include Income, Median Age, % 65+, % Bachelor+, % Renters, Year Structure Built, Avg. No. Bedrooms, Median Home Value, Rurality, % Dem. Voters, % African American, % Hispanic, % Asian, % Other Race, % Tree-to-Land Area, Solar Radiation

Table C5 Full Model

| Predictors | PV Count Per Household | | PV-to-Roof Area | |
|---|---|---|---|---|
| | Model 31 Coef. (SE) | Model 32 Coef. (SE) | Model 33 Coef. (SE) | Model 34 Coef. (SE) |
| % Tree-to-Land Area | .389 (.545) | 2.057 (.878)* | −.550 (.115)*** | −.476 (.175)** |
| % Tree-to-Land Area$^2$ | −51.098 (17.448)* | −145.512 (36.07)** | 2.666 (3.669) | 1.140 (7.177) |
| Solar Radiation | .068 (.095) | .207 (.224) | .005 (.020) | −.054 (.045) |
| Solar Radiation$^2$ | −.006 (.008) | −.018 (.019) | .001 (.002) | .004 (.004) |
| Median HH Income | .001 (.001) | .001 (.001)** | .001 (.001) | −.001 (.001) |
| Median HH Income$^2$ | −.001 (.001) | −.001 (.001)* | .001 (.001)* | .001 (.001) |
| Median Age | −.001 (.001) | .001 (.001) | −.001 (.001)*** | −.001 (.001)** |
| % 65+ | .001 (.017) | −.017 (.021) | .001 (.004) | .003 (.004) |
| % Bachelor+ | .032 (.018)† | −.023 (.024) | .004 (.004) | −.003 (.005) |
| % Renters | −.026 (.012)* | −.039 (.015)** | .003 (.003) | .001 (.003) |
| Year Structure Built | .120 (.017)*** | .089 (.022)*** | .017 (.004)*** | .010 (.004)* |
| Year Structure Built$^2$ | −.001 (.001)*** | −.001 (.001)*** | −.001 (.001)*** | −.001 (.001)* |
| Avg. No. Bedrooms | .002 (.005) | .001 (.006) | −.007 (.001)*** | −.009 (.001)*** |
| Median Home Value | .001 (.001)*** | .001 (.001)*** | .001 (.001)*** | .001 (.001)*** |
| Rurality | −.004 (.001)** | .004 (.003) | −.001 (.001)*** | −.001 (.001) |
| % Dem. Voters | .082 (.016)*** | .166 (.029)*** | .010 (.003)** | .013 (.006)* |
| % African American | −.019 (.023) | −.041 (.026) | .017 (.005)*** | .007 (.005) |
| % Hispanic | −.0145 (.013) | −.012 (.018) | −.004 (.003) | −.004 (.004) |
| % Asian | .205 (.036)*** | .140 (.041)*** | .017 (.008)* | .005 (.009) |
| % Other Race | .076 (.036)* | .068 (.052) | 011 (.007) | .017 (.012) |
| Transmission Volt. | −.004 (.001)*** | −.001 (.003) | .001 (.001) | .001 (.001) |
| Transmission Length | .001 (.001) | .001 (.002) | −.001 (.001) | −.001 (.001) |
| Muni. Utilities | −.055 (.006)*** | −.061 (.009)** | −.004 (.001)*** | −.004 (.002)† |
| Rural Co-ops. | −.046 (.005)*** | −.013 (.011) | −.007 (.001)*** | −.006 (.002)** |
| Resident. Elec. Rate | −.921 (.142)*** | .429 (.351) | .013 (.030) | .235 (.070)*** |
| Commerc. Elec. Rate | .975 (.167)*** | −1.246 (.644)† | .015 (.035) | −.351 (.128)** |
| Solar Mandate | | .015 (.009)† | | .005 (.002)** |
| Net Metering | | .033 (.011)** | | .006 (.001)* |
| SolSmart Awardee | | −.004 (.007) | | −.001 (.001) |
| Online Permitting | | -.001 (.006)* | | −.001 (.001) |
| Sameday Inp. Per. | | −.008 (.008) | | −.001 (.002) |
| Per. & Pre–Inst. Days | | −.002 (.001)*** | | −.001 (.001)*** |
| Drought Risk | −.001 (.001) | −.011 (.002)*** | .001 (.001) | −.001 (.001)** |
| Wildfire Risk | −.001 (.001)*** | .001 (.001)† | −.001 (.001) | .001 (.001)*** |
| Hail Risk | −.001 (.001)*** | −.001 (.001)** | .001 (.001) | .001 (.001) |
| Winter Weather Risk | .001 (.001)* | −.001 (.001) | .001 (.001)*** | .001 (.001) |
| Strong Wind Risk | .002 (.001)*** | −.004 (.001)*** | .001 (.001) | .001 (.001) |
| Tornado Risk | −.002 (.001)*** | −.003 (.001)*** | −.001 (.001)*** | .001 (.001) |
| % Below Poverty | −.001 (.021) | −.007 (.027) | .014 (.004)** | .015 (.005) |
| % Disability | .001 (.001) | −.001 (.001) | −.001 (.001)† | −.001 (.001)* |
| % Single Parent | .001 (.001) | .001 (.001) | -.001 (.001) | −.001 (.001) |
| % Limit. English | −.001 (.001) | −.001 (.001) | .001 (.001) | .001 (.001) |
| % 10+ Unit Housing | −.001 (.001)*** | −.001 (.001)*** | .001 (.001)*** | .001 (.001)*** |
| % Mobile Homes | −.001 (.001) | .001 (.001) | .001 (.001)*** | −.001 (.001)*** |
| % Ppl. > Rooms | −.001 (.001) | .001 (.001) | .001 (.001)** | .001 (.001)*** |
| % No Vehicle | −.001 (.001) | −.001 (.001) | .001 (.001) | −.001 (.001) |
| % Unemployed | .001 (.001) | .001 (.001) | .001 (.001)*** | .001 (.001)* |
| No. of block groups | 3441 | 2328 | 3441 | 2328 |
| $R^2$ | .262 | .329 | .271 | .348 |

*** $p < 0.001$, ** $p < 0.01$, * $p < 0.05$, † $p < 0.1$. Robust Standard Errors are in parentheses.

# APPENDIX D. A trial run illustrating accuracy obtained for identifying roofs

```
IoU metric: bbox
 Average Precision  (AP) @[ IoU=0.50:0.95 | area=   all | maxDets=100 ] = 0.661
 Average Precision  (AP) @[ IoU=0.50      | area=   all | maxDets=100 ] = 0.962
 Average Precision  (AP) @[ IoU=0.75      | area=   all | maxDets=100 ] = 0.773
 Average Precision  (AP) @[ IoU=0.50:0.95 | area= small | maxDets=100 ] = 0.091
 Average Precision  (AP) @[ IoU=0.50:0.95 | area=medium | maxDets=100 ] = 0.602
 Average Precision  (AP) @[ IoU=0.50:0.95 | area= large | maxDets=100 ] = 0.742
 Average Recall     (AR) @[ IoU=0.50:0.95 | area=   all | maxDets=  1 ] = 0.113
 Average Recall     (AR) @[ IoU=0.50:0.95 | area=   all | maxDets= 10 ] = 0.718
 Average Recall     (AR) @[ IoU=0.50:0.95 | area=   all | maxDets=100 ] = 0.725
 Average Recall     (AR) @[ IoU=0.50:0.95 | area= small | maxDets=100 ] = 0.371
 Average Recall     (AR) @[ IoU=0.50:0.95 | area=medium | maxDets=100 ] = 0.665
 Average Recall     (AR) @[ IoU=0.50:0.95 | area= large | maxDets=100 ] = 0.801
```

**Competing interests**
The authors declare that they have no competing interests.

**Funding**
This research was supported by the Presidential Initiative on Urban and Place-Based Research and the Education Through Undergraduate Research and Creative Activities program at the University of Colorado Denver.

**Author's contributions**
SYK conceived the idea, secured the funding, collected data, performed data analysis, and wrote the manuscript. KG designed and implemented the computer vision model and wrote the manuscript. CS collected data and wrote the manuscript. RO implemented computer vision model and performed ML model experiments. All authors read and approved the manuscript.

**Abbreviations**
CDC, Centers for Disease Control and Prevention; DER, Distributed Energy Resources; dGEN, Distributed Generation Market Demand; DNI, Direct Normal Irradiance; FEMA, Federal Emergency Management Agency; FIS, Feature Importance Score; GIS, Geographical Information Systems; HH, Household; ML, Machine Learning; NEM, Net Metering; NREL, National Renewable Energy Laboratory; PII, Permitting, Inspection, and Interconnection; PV, Photovoltaic; RPS, Renewable Energy Portfolio Standards; SHAP, SHapley Additive exPlanations; SVI, Social Vulnerability Index.

**Availability of data and materials**
Datasets and Python code used in data analysis and visualization in this research will be available on Git Large File Storage and GitHub upon publication.

**Author details**
[1]College of Engineering, Design and Computing, University of Colorado Denver, 1200 Larimer St., Denver, CO USA, 80204. [2]School of Public Affairs, University of Colorado Denver, 1380 Lawrence St., Denver, CO USA, 80204. [3]Computer Science, University of Colorado Boulder, 1111 Engineering Dr., Boulder, CO USA, 80309. [4]Physics, University of Colorado Boulder, 2000 Colorado Ave., Boulder, CO USA, 80309. [5]Computer Science, University of Colorado Denver, 1380 Lawrence St., Suite 3034, Denver, CO USA, 80204. [6]RadiaSoft, 6525 Gunpark Dr, Suite 370-411, Boulder, CO USA, 80301.